\documentclass[prd,aps,twocolumn,amsfonts,showpacs,superscriptaddress,preprintnumbers,footinbib]{revtex4-2}

\usepackage{mathtools}                                                                                                                                
\usepackage{graphicx,color}
\usepackage{dcolumn}
\usepackage{bm}

\usepackage[colorlinks=true, urlcolor=blue, linkcolor=blue, citecolor=blue, hyperindex=true, linktocpage=true]{hyperref}

\newcommand{\mn}{{\mu\nu}}

\newcommand{\sig}{{\hat{\sigma}}}

\newcommand{\tone}{\tau_R}
\newcommand{\ttwo}{\tau_C}

\begin{document}

\title{Towards the BBGKY hierarchy: a scheme beyond the Boltzmann equation and application to a weakly confined QCD gas}

\author{Sa\v{s}o Grozdanov}
\affiliation{Higgs Centre for Theoretical Physics, University of Edinburgh, Edinburgh, EH8 9YL, Scotland}
\affiliation{Faculty of Mathematics and Physics, University of Ljubljana, Jadranska ulica 19, SI-1000 Ljubljana, Slovenia}
 
\author{Alexander Soloviev}
\affiliation{Faculty of Mathematics and Physics, University of Ljubljana, Jadranska ulica 19, SI-1000 Ljubljana, Slovenia}

\begin{abstract}
In classical kinetic theory, the BBGKY hierarchy is an infinite chain of integro-differential equations that describes the full time-reversal-invariant (Liouville) system of interacting (quasi)-particles in terms of $N$-particle distribution functions. In this work, instead of truncating the hierarchy at the lowest level, as is done by the Boltzmann equation, we develop a scheme similar to the relaxation time approximation that is in principle able to account for the entire chain of equations. We then explicitly investigate its truncation at the second level of the BBGKY hierarchy and, within this scheme, study the spectra of conserved operator correlation functions in a gas of weakly confined hadrons. We also discuss how these higher levels account for parts of the operator spectra `deeper in the ultra-violet regime' and compare them to known results derived from the holographic duality.   
\end{abstract}

\maketitle

\textbf{Introduction.---}Kinetic theory is one of cornerstones of theoretical physics, with applications going far beyond Boltzmann's first conception, including plasma physics and fusion \cite{plasma-fusion}, high-energy particle physics experiments like in heavy ion collisions \cite{Busza:2018rrf} and cosmology \cite{galaxy-scattering}. It is described concisely in the so-called BBGKY (Bogoliubov-Born-Green-Kirkwood-Yvon) hierarchy \cite{yvon,bogoliubov,kirkwood,borngreen}, which represents an exact treatment of an $N$-particle system. 

Although theoretically complete, the BBGKY hierarchy is impractical for computations and a number of approximations must be made, most often truncating it at the first level of the hierarchy (the Boltzmann equation) and choosing some form of the collision operator \footnote{We also note some recent developments in using the BBGKY in nearly integrable systems \cite{biagetti2024generalisedbbgkyhierarchynearintegrable}.}.  The simplest truncation is the relaxation time approximation (RTA), also known as the Bhatnagar-Gross-Krook approximation \cite{bgk} with the Anderson-Witting model as its relativistic version \cite{ANDERSON1974466}. It is a physically-motivated and analytically solvable near-equilibrium scheme in which the collision integral is assumed to have a single dominant (late-time) relaxation time $\tone$. Among its other applications, it has been used to compute correlation functions of conserved operators \cite{Romatschke:2015gic,Bajec:2024jez}. The approximation scheme was generalized in \cite{Kurkela:2017xis} to study the effects of momentum-dependent relaxation time. On the other hand, going beyond the RTA, the `AMY' kinetic theory \cite{Arnold:2002zm} has been used to describes weakly coupled QCD at high temperature. Other recent solutions of the Boltzmann equation include \cite{Florkowski:2014sfa,Denicol:2014xca,Denicol:2022bsq}.

The RTA Boltzmann equation in which multiparticle correlations are neglected can work well when all other `microscopic' timescales in the problem are shorter than $\tone$. One such timescale is the correlation time, $\ttwo$. Typically, for dilute weakly coupled gases, $\tone\gg \ttwo$, and one can disregard the higher levels of the BBGKY hierarchy. However, these assumptions seem unsatisfactory for a description of dense and strongly coupled systems broadly accessible in condensed matter physics and in the context of heavy ion collisions, in particular, if one is interested in earlier timescales, like in prescaling and approaches to pre-equilibrium \cite{Berges:2020fwq}. This regime occurs well before hydrodynamic attractor behavior takes over \cite{Soloviev:2021lhs,Jankowski:2023fdz}. 

In this work, we develop a scheme to access higher levels of the BBGKY hierarchy. We apply it to the study of the BBGKY hierarchy to second level in the correlation time approximation (CTA), with the second-level collision operator approximated like in the RTA. Unlike in previous works \cite{bonitz_1996}, the resulting linearized equations can be solved and correlation functions of conserved operators that control late-time physics can be determined analytically. The second aim of this work is to apply the extended hierarchy to an example of a physical system: a QCD gas just below the confining temperature $\sim155$ MeV. In this case, the particles feel a weak, long-range confining force. All our explicit calculations are done in such a setting that combines standard RTA techniques with a long-range potential that introduces sensitivity to nontrivial two-particle correlations explicitly coupling the first two levels of the BBGKY hierarchy. 

\textbf{The BBGKY hierarchy.---}The Liouville equation describes the flow of invariant volume of phase space describing an isolated system of $N$ particles \cite{liboff-book}.~It expresses the evolution of the $N$-particle distribution function $f_N=f_N ({\bf r}_1, \ldots {\bf r}_N,  {\bf p}_1,\ldots,{\bf p}_N;t)$:
\begin{align}\label{Liu}
\partial_t f_N-\{f_N, H_N \} = 0,
\end{align}
where the Poisson bracket is defined as  
\begin{align}
\{A, B \}=\frac{\partial A}{\partial {\bf r}}\cdot \frac{\partial B}{\partial {\bf p}}-\frac{\partial B}{\partial {\bf r}}\cdot \frac{\partial A}{\partial {\bf p}}.
\end{align}
The $n$-particle Hamiltonian for a collection of particles weakly interacting via a central force is given by
\begin{align}\label{ham}
H_n=\sum_{i=1}^n \left( \frac{{{\bf p}}_i^2}{2m}+V({\bf r}_i) \right) +\sum_{i<j<n} U({\bf r}_i-{\bf r}_j).
\end{align}
$V$ and $U$ are external and interparticle potentials, respectively. Since one can typically only keep track of a few correlations at a time, next, we define the reduced particle distribution functions given in terms of $f_N$ by
\begin{align}\label{nlevels}
&f_n({\bf r}_1, \ldots {\bf r}_n,  {\bf p}_1,\ldots,{\bf p}_n;t)=\frac{N!}{(N-n)!}  \\
&\times \int \Pi_{i=n+1}^N d^3 r_i d^3 p_i f_N({\bf r}_1, \ldots {\bf r}_N,  {\bf p}_1,\ldots,{\bf p}_N;t) .\nonumber
\end{align} 

It is well known that the evolution of the $N$-particle distribution function \eqref{Liu} is equivalently represented as a tower of $N$ equations --- the BBGKY hierarchy. In particular, the classical evolution of the $n$-point distribution function is given by
\begin{align}\label{fn-evolution}
\partial_t f_n -\{f_n, H_n \}=\sum_{i=1}^n \int d^3 r_{n+1}d^3 p_{n+1} \frac{\partial U}{\partial {\bf r}}\cdot \frac{\partial f_{n+1}}{\partial {\bf p}} .
\end{align}
At each level $n$, the right-hand side, or the collision kernel, depends on $f_{n+1}$. We denote it by $C[f_{n+1}]$. Hence, to turn the infinite chain of equations to a finite closed system, we clearly require some truncation scheme. 

The one-particle distribution $f_1$ plays a special role in this formalism in that it allows us to compute conserved currents, such as a conserved number current $J^\mu$ and the energy-momentum tensor $T^{\mu\nu}$ via
\begin{align}\label{current-emt}
    J^\mu = \int \frac{d^3 p}{(2\pi)^3} \frac{p^\mu}{p^0} f_1, \quad
    T^\mn = \int \frac{d^3 p}{(2\pi)^3} \frac{p^\mu p^\nu}{p^0} f_1,
\end{align}
where we used the relativistic notation in 4$d$ spacetime denoted by Greek indices. 

\textbf{Beyond the relaxation time approximation.---}To compute $f_1=f_1(\textbf{r},\textbf{p};t)$, which couples to $f_2=f_2(\textbf{r}_1,\textbf{p}_1,\textbf{r}_2,\textbf{p}_2;t)$ through $C[f_2]$, it is standard to follow Boltzmann's reasoning. In particular, we assume the gas to be dilute (suppressing higher $f_n$) and the so-called molecular chaos ansatz factorizing $f_2 = f_1 f_1$. This decouples higher levels of the hierarchy, producing a single closed integro-differential (Boltzmann) equation for $f_1$. The latter assumption also introduces an arrow of time and a growing entropy. In a further simplification, one may assume a system linearized around equilibrium $f_1 = f_1^{\rm eq}$ with a single eigenmode that dominates late-time relaxation. This reduces the Boltzmann equation to its RTA form:
\begin{align}
    \partial_t f_1+{\bf v}\cdot \frac{\partial }{\partial {\bf r}} f_1 +{\bf F}\cdot\frac{\partial}{\partial {\bf p}} f_1  \approx C_{\rm RTA} = -\frac{f_1-f_1^{\rm eq}}{\tone},
\end{align}
where ${\bf v}={\bf p}/p^0$ and ${\bf F}$ is an external force. 

We go beyond the RTA Boltzmann equation and develop a scheme that accounts for correlations governed by higher levels of the BBGKY hierarchy. In particular, we decompose the two-particle distribution function into uncorrelated and correlated pieces \cite{liboff-book}, 
\begin{align}
    f_2=f_1 f_1+g_{12},
\end{align}
and approximate the uncorrelated contribution $f_1f_1$ to the lowest level of the hierarchy using standard RTA. On the other hand, we keep the evolution of the correlated piece, $g_{12}$, in $C[f_2]$ exact. In other words,
\begin{align}\label{collision-decomp}
    C[f_2]&= C[f_1 f_1]+ C[g_{12}] \approx-\frac{f_1-f_1^{\rm eq}}{\tone} 
    + C[g_{12}].
\end{align}
We note that such a decomposition is possible because the collision kernel acts linearly on $f_2$ (cf.~Eq.~\eqref{fn-evolution}). 

As in standard RTA, we need to enforce the energy-momentum and current conservation by adapting the RTA matching conditions (see e.g.~\cite{Romatschke:2015gic,Bajec:2024jez}) to 
\begin{align}
    u_\mu \left(T^\mn - T^\mn_{\rm eq} +\tone T^\mn_{\rm corr}\right)&=0, \\
     u_\mu \left(J^\mu - J^\mu_{\rm eq} +\tone J^\mu_{\rm corr} \right)&=0.
\end{align}
The terms with a subscript ``$\rm corr$'' denote the integral over the correlated piece, e.g.
\begin{align}
    T^\mn_{\rm corr}=\int\frac{d^3p}{(2\pi)^3} \frac{p^\mu p^\nu}{p^0} C[g_{12}].
\end{align}

We can now sketch how the approximation scheme works at the $n^{\rm th}$ level of the hierarchy. We decompose the $n$-particle distribution function into a sum of products of lower level particle distribution functions
\begin{align}
    f_n = (f_1)^n +\sum g_{12}f_1^{n-2}+\ldots + g_{1\ldots n},
\end{align}
where the sum runs over particle numbers and $g_{1\ldots n}$ represents the fully correlated piece, not composed of lower particle distribution functions. For example, $f_3=f_1(1)f_1(2)f_1(3) + f_1{(1)}g{(2,3)}+f_1(2) g(1,3) + f_1(3) g(1,2) +g{(1,2,3)}$, where $f_1(M) = f_1({\bf r}_M, {\bf p}_M; t)$, and similarly for $g$. Using lower level equations and approximating the uncorrelated piece using an independent relaxation time at each level, $\tau_n$, the evolution equation of the $n$-particle distribution function is then 
\begin{align}\label{hevolution}
\partial_t g_{1 \ldots n} -\{g_{1 \ldots n}, H_n \}
    =-\frac{g_{1\ldots n}-g_{1\ldots n}^{\rm eq}}{\tau_n} + C[g_{1\ldots n+1}].
\end{align}
When working to level $n$, we neglect all higher orders in the hierarchy (setting $C[g_{1\ldots n+1}] =0 $), thus closing the set of equations. We refer to this procedure as the correlation time approximation (CTA). Note that in equilibrium, $f_n^{\rm eq}=0$ for $n \geq 2$, from which $g_{1\ldots n}^{\rm eq}$ can be deduced. It is straightforward to see that the CTA still leads to positive entropy production (the H-theorem). 

\textbf{Two-level BBGKY hierarchy in the CTA applied to a QCD gas.---}We now turn our attention to an example of the two-level BBGKY analysis. We consider a QCD gas at energies just below the confinement phase transition. The interparticle potential is composed of two pieces: the standard scattering short-range potential and a long-range weak linear confining QCD potential: 
\begin{align}\label{cornell}
    U(r)= U_{\rm S}(r) + U_{\rm L}(r).
\end{align}
An example is the Cornell potential \cite{Deur:2016tte} in which $U_{\rm S}(r) = \alpha / r$ is the Coulomb potential and $U_{\rm L}(r) = \sigma r$, where $\sigma$ is the string tension. The  linear QCD potential $U_{\rm L}$ is what now allows us to introduce non-trivial correlations that occur beyond from the `local' scatterings induced by $U_{\rm S}$ in standard kinetic theory treatments. The scales at which $U_{\rm L}$ should become relevant are greater than $1$ fm, whereas the Coulomb potential dominates on smaller scales \cite{Bali:1992ab}. To access long-range transport properties of such systems, we further simplify \eqref{collision-decomp} by neglecting the effects of $U_{\rm S}$ on the correlations encoded in $g_{12}$ (cf.~Eq.~\eqref{fn-evolution}). This is because we expect that the short-range interactions prevent the build up of correlations. A different point of view on these approximations is as the simplest possible way to couple the first two levels of the hierarchy, keeping collision kernels at each level RTA-like. 

We are left with the following (Fourier transformed) term linking the first two levels of the hierarchy 
\begin{align}
&\int \frac{d^3 r_1}{(2\pi)^3} e^{-i \textbf{k}\cdot \textbf{r}_1}\int d^3 r_2 d^3 p_2  \frac{\partial U_{\rm L}({\bf r}_1-{\bf r}_2)}{\partial \textbf{r}_1}\cdot \frac{\partial g_{12}}{\partial \textbf{p}_1}  \\
   &= (2\pi)^3 i \!  \int d^3 p_2 d^3 Q  U_{\rm L}(\textbf{Q})  \textbf{Q}\cdot \frac{\partial g_{12}(\textbf{k}-\textbf{Q},\textbf{p}_1,\textbf{Q},\textbf{p}_2;t)}{\partial \textbf{p}_1}, \nonumber
   \end{align}
where $U_{\rm L} (\textbf{Q}) = -i \sig  \delta^{\prime}(\textbf{Q})$, $\sig = \sigma V$, and we take the characteristic volume $V\sim R^3$ to be determined by the size of the quark-gluon plasma droplet with $R\sim 1$ fm \cite{Busza:2018rrf}.
Finally, we are interested in the regime of small $\sigma/\Lambda_{\rm QCD}^2$ in which the linear coupling can be treated perturbatively also as compared to $V$. This occurs at temperatures below the confining temperature of roughly $155$ MeV, where some quarks begin to form confining pairs. Indeed, as confirmed by lattice calculations, in this regime, $\sigma/\Lambda_{\rm QCD}^2$ is small \cite{Kaczmarek:1999mm,Bicudo:2010hg}. Eventually, the system cools enough until the string tension approaches its usually quoted value of $\sigma=0.18$ GeV$^2$ for the QCD quark-antiquark potential \cite{Bali:2000gf,Deur:2016tte} (for a holographic model, see~\cite{Trawinski:2014msa}). 

With these physical considerations in hand, the resulting equations coupling levels one and two of the BBGKY hierarchy in the CTA are 
\begin{align}\label{level1}
    &\left(\partial_t+\textbf{v}_1\cdot \frac{\partial}{ \partial \textbf{r}_1}
    +{\textbf{F}}_1\cdot \frac{\partial}{ \partial \textbf{p}_1}\right)f_1=-\frac{f_1-f_1^{\rm eq}}{\tone}\nonumber\\
    &-\int d^3 r_2 d^3 p_2  \frac{\partial U_{\rm L}({\bf r}_1-{\bf r}_2)}{\partial \textbf{r}_1}\cdot \frac{\partial g_{12}}{\partial \textbf{p}_1},\\
    &\left(\partial_t+\textbf{v}_a\cdot \frac{\partial}{ \partial \textbf{r}_a}
    +{\textbf{F}}_a \cdot \frac{\partial}{ \partial \textbf{p}_a}\right)g_{12}=-\frac{g_{12}-g_{12}^{\rm eq}}{\ttwo},\label{level2}
\end{align}
where we sum over the index $a=1,2$. The equations can be solved by linearizing about equilibrium and going into Fourier space. We write
\begin{align}
f_1(\textbf{r}_1,\textbf{p}_1;t)&=f_1^0(\textbf{p}_1)+\delta f_1(\textbf{r}_1,\textbf{p}_1;t),\\
g_{12}(\textbf{r}_1,\textbf{p}_1,\textbf{r}_2,\textbf{p}_2;t)
&=g_{12}^0(\textbf{p}_1,\textbf{p}_2)+\delta g_{12}(\textbf{r}_1,\textbf{p}_1,\textbf{r}_2,\textbf{p}_2;t),
\end{align}
with the equilibrium values of the massless gas given in terms of the Maxwell-Boltzmann distribution
\begin{align}\label{dist-eq}
    f_1^0&=f_1^{\rm eq}=e^{-\frac{p-\mu_0}{T_0}},\\
    g_{12}^0&=g_{12}^{\rm eq}=-f_1^0(\textbf{p}_1)f_1^0(\textbf{p}_2) .
\end{align}
Note the sign choice in the above guarantees that in equilibrium, $f_2^{\rm eq}=f_1^{\rm eq}f_1^{\rm eq}+g_{12}^{\rm eq}=0$, as discussed above.

\textbf{Charge transport.---}To study charge transport, we turn on an external electromagnetic gauge field $A^\mu$, which gives an external field $\textbf{E}$ and a force $\textbf{F}=q\textbf{E}$ (we set $q=1$). The variation of the equilibrium distribution function is then
\begin{align}
    \delta f_1^{\rm eq}&= \frac{f^0_1(\textbf{p}_1)}{T_0} \delta \mu(t,x_1),\\
    \delta g_{12}^{\rm eq}&=\frac{g_{12}^0(\textbf{p}_1,\textbf{p}_2)}{T_0} \left[\delta \mu(t,x_1)+\delta \mu(t,x_2)\right].
\end{align}
The solution to \eqref{level2} in the presence of ${\bf E}$ is then
\begin{align}
    &\delta g_{12}(\textbf{k}_1,\textbf{k}_2,\textbf{p}_1,\textbf{p}_2;\omega)\\
    &=\frac{g^0_{12}}{T_0}\frac{\textbf{v}_a\cdot \textbf{E}_a+ \left[ \delta \mu(\omega,k_1)+\delta \mu (\omega,k_2) \right] /\ttwo}{-i\omega +i \textbf{v}_a\cdot \textbf{k}_a+1/\ttwo} .  \nonumber
\end{align}
Note that ${\bf E}_1={\bf E}(\omega,\textbf{k}_1)$ and ${\bf E}_2={\bf E}(\omega,\textbf{k}_2)$. Using the solution to the second level, the first level reads
\begin{align}\label{level1}
    &\left(-i\omega + i \textbf{v}\cdot \textbf{k}+\frac{1}{\tone}\right) \delta f =\frac{f_1^0}{T_0}\left(\textbf{v}\cdot \textbf{E}+\frac{\delta \mu}{\tone}\right)\\
    &+\sig\int d^3 p_2   \frac{\partial  g_{12} ( \textbf{k},0,\textbf{p}_1,\textbf{p}_2;\omega)}{\partial p_1 }.\nonumber
\end{align}
The change to the current is given by \eqref{current-emt}, 
\begin{align}
    \delta J^\mu = \int \frac{d^3 p_1}{(2\pi)^3 p^0} p^\mu\delta f,
\end{align}
where $\delta J^0=\delta  n= \chi \delta \mu$ and the static susceptibility is
\begin{align}
    \chi =\int dp \frac{p^2}{2\pi^2 T_0}f_0.
\end{align}
Next, we need to solve self-consistently for $\delta n$, i.e., we need to impose the RTA matching conditions. Note that unlike in \cite{Romatschke:2015gic,Bajec:2024jez}, we need to additionally solve for the zero momentum contribution to the number density, $\delta n(\omega,0)$. We find $\delta n (\omega,0)=0.$ We then compute the correlators via the variational principle, namely,
\begin{align}
G^{\mu,\nu}_J=\frac{\delta J^\mu}{\delta A_\nu}.   
\end{align}
Explicit expressions of the correlators can be found in Appendix~\ref{app:correlators}. All correlators we compute are expressed to leading order in $\sigma$ as
\begin{align}
    G^{\mu,\nu}_J(\omega,k)&= \frac{A(\omega,k)+\sig B(\omega,k)+\mathcal{O}(\sig^2)}{C(\omega,k)+\sig D(\omega,k)+\mathcal{O}(\sig^2)},
\end{align}
where the $\sig\rightarrow 0$ limit corresponds to the results in \cite{Romatschke:2015gic}. 

\begin{figure*}
    \centering
\includegraphics[width=0.33\linewidth]{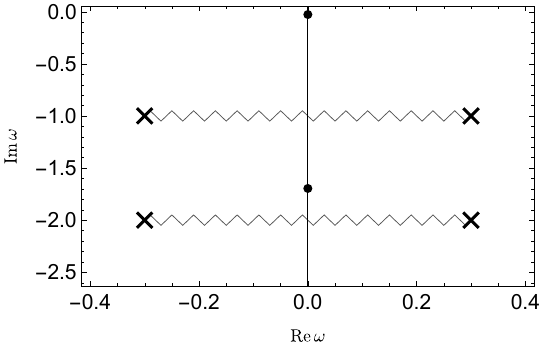}\includegraphics[width=0.33\linewidth]{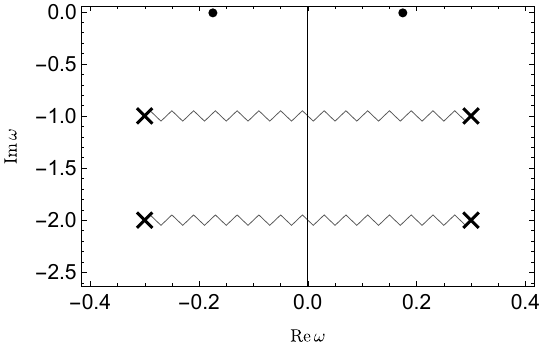}\includegraphics[width=0.33\linewidth]{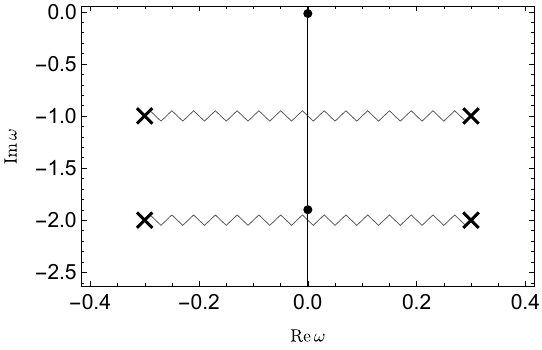}
    \caption{Left: The analytic structure of $G^{0,0}_{J}$ in the complex frequency $\omega$ plane. Parameters for the plot are $k=0.3$, $\chi=1$, $2\ttwo=\tone=1$ and $\hat\sigma=0.18
$. Middle: The analytic structure of the longitudinal (sound channel) $G^{00,00}_T$ in the complex frequency $\omega$ plane, with the same parameters as in the left panel. Right: The analytic structure of the transverse (shear channel) $G^{01,01}_T$ in the complex frequency $\omega$ plane for $\hat\sigma=1.18$. In all figures, dots depict poles and crosses represent branch points.}
\label{fig:correlators}
\end{figure*}

Focusing first on the analytic structure of the longitudinal (density) correlator $G^{0,0}_J$, we explicitly see how the inclusion of the second hierarchy level allows us to go beyond the Boltzmann RTA results (see \cite{Romatschke:2015gic,Bajec:2024jez}). Instead of a single branch cut, we have two logarithmic branch cuts, one corresponding to each `relaxation time' with branch points at 
\begin{align}\label{branch1}
    \omega &= \pm k -\frac{i}{\tone}, \quad 
    \omega= \pm k -\frac{i}{\ttwo}.
\end{align}
The branch cuts correspond to collective (continuum of) excitations of the gas \cite{Kurkela:2017xis}, which decay on a timescale slower than exponentially (as would be the case for poles). This is a typical structure seen in, e.g.~hard thermal loop calculations \cite{Blaizot:2001nr}.
The correlator also has the (hydrodynamic) diffusive pole present in RTA and a new gapped pole located between the branch cuts. Their dispersion relations to $\mathcal{O}(\sig)$  are
\begin{align}
    \omega&=
    -i D k^2 + \mathcal{O}(k^3), \nonumber \\
    \omega&=-\frac{i}{\ttwo} \left( 1 - \hat\sigma \frac{\chi \tone}{ T_0 } \right) - i k^2\biggr[ \frac{T_0 \ttwo}{3   \sig  \tone \chi } -\frac{ \ttwo^2}{3 (\tone-\ttwo)} \nonumber \\
    &+\frac{  \sig   \chi  \tone \ttwo^2 \left(2 \tone -  \ttwo\right)}{3T_0 (\tone-\ttwo)^2} \biggr]+ \mathcal{O}(k^3),
\end{align}
where $D$ is charge diffusion coefficient. The gapped pole emerges due to the second level of the hierarchy characterized by $\ttwo$. Explicitly, $D$ is given by
\begin{align}
    D=
    \frac{\tone}{3}\left( 1- \sig\frac{\chi \ttwo }{T_0}\right) +\mathcal{O}(\sig^2),
\end{align}
The analytic structure of the longitudinal (density) correlator $G^{0,0}_J$ is shown in the left panel of Fig.~\ref{fig:correlators}. 
The transverse current-current $G^{1,1}_J$, which contains no hydrodynamic poles, exhibits only both branch cuts. 

\textbf{Energy-momentum transport.---}To study the energy-momentum transport, we perturb the system around the Minkowski background, $\eta_\mn,$ via a perturbation, $g_\mn=\eta_\mn + \delta g_\mn,$ which induces a change in the macroscopic quantities, namely the temperature $T=T_0+\delta T$ and four velocity $u^\mu=(1,0,0,0)+\delta u^\mu$, which is normalized $u^\mu u_\mu=-1$. The external force term is now
\begin{align}
    F^i_a= \Gamma^i_{a,\alpha \beta}\frac{p^\alpha p^\beta}{p^0},
\end{align}
where $\Gamma^\mu_{\alpha \beta}=\Gamma^\mu_{\alpha \beta}\left[\delta g_\mn\right]$ is  the Christoffel symbol. Note that $\Gamma^i_{1,\alpha \beta}=\Gamma^i_{\alpha \beta}(\omega,\textbf{k}_1)$ and $\Gamma^i_{2,\alpha \beta}=\Gamma^i_{\alpha \beta}(\omega,\textbf{k}_2)$.
The evolution of the one-particle distribution function is then
\begin{align}
    &\left(-i\omega + i \textbf{k}\cdot\textbf{v}\right)\delta f
    +\frac{f_0}{T_0}  \Gamma^0_{\alpha \beta} p^0 v^\alpha v^\beta   \\
    &= -\frac{\delta f- \delta f^{\rm eq}}{\tau_R}+\sig\int d^3 p_2  
    \partial_{p_1}{g_{12}}(\omega,\textbf{k},0,\textbf{p}_1,\textbf{p}_2). \nonumber
\end{align}
The change to the equilibrium statistical distributions is
\begin{align}
    \delta f^{\rm eq}&=f^0 \frac{p^0}{T_0}\left(\textbf{v}\cdot\delta \textbf{u} +\frac{\delta T}{T_0}\right),\\
    \delta g_{12}^{\rm eq}&=g^0_{12}\frac{p^0_a}{T_0}\left(\textbf{v}_a\cdot\delta \textbf{u}_a +\frac{\delta T_a}{T_0}\right),
\end{align}
which allows us to determine the energy-momentum tensor via \eqref{current-emt}. The $T^{\mu\nu}$ correlators are then given by
\begin{align}
G^{\mn,\alpha\beta}_T=\frac{\delta T^\mn}{\delta \delta g_{\alpha \beta}}   
\end{align}
and its explicit expressions are given in Appendix~\ref{app:correlators}. 

The analytic structure of the $T^{\mu\nu}$ correlators again exhibits the same two branch cuts with branch points \eqref{branch1}. Moreover, the longitudinal $G^{00,00}_T$ correlator, shown in the middle panel of Fig.~\ref{fig:correlators}, contains the sound mode:
\begin{align}
    \omega=\pm\frac{1}{\sqrt{3}}k- i \left(\frac{2 \tone}{15}-\sig\frac{\tone \ttwo T_0^2}{10\pi^2}\right)k^2  +\mathcal{O}(k^3).
\end{align}
Curiously, the energy correlator $G^{00,00}_T$ seemingly has an additional pole but with zero residue for all values of parameters and for all $\omega$ and $k$ to leading order in $\sig$.

The transverse (shear) $G^{01,01}_T$ correlator is shown in the right panel of Fig.~\ref{fig:correlators}, which in addition to the branch cuts, also has two modes (momentum diffusion and a gapped mode) with dispersion relations to $\mathcal{O}(\sig)$:
\begin{align}
    \omega&=-\frac{i \tau_R}{5}\left(1-\sig \frac{3     \ttwo T_0^2}{4 \pi ^2} \right)k^2+\mathcal{O}(k^3),\\
    \omega&=-\frac{i}{\ttwo}\left(1-\sig \frac{3   T_0^2 \tone}{4 \pi ^2
   }\right)-i k^2 \biggr[
   \frac{4  \pi ^2 \ttwo}{15 \sig  T_0^2 \tone}\\
   &
   -\frac{ \ttwo^2}{5 (\tone-\ttwo)}
   +\frac{3     \sig  T_0^2 \tone \ttwo^2 (2
   \tone-\ttwo)}{20 \pi ^2 (\tone-\ttwo)^2}\biggr] +\mathcal{O}(k^3).\nonumber
\end{align}

Finally, the analytic structure of the tensor channel correlator $G_T^{12,12}$ only exhibits the two branch cuts and no poles. 

\textbf{From weak to strong coupling and from IR to the UV.---}It is interesting to explore the ratio of the shear viscosity $\eta$ to entropy density  $s=(\varepsilon_0+P_0)/T_0$, which controls first-order hydrodynamic momentum diffusion and sound attenuation. It can be computed from any of the $G^{\mu\nu,\rho\sigma}_T$ channels. It is given by
\begin{align}\label{eta-s}
    \frac{\eta}{s}=\frac{\tau_R T_0}{5}\left(1
    -\sig \frac{3\ttwo T_0^2}{4\pi^2}+\mathcal{O}(\sig^2)\right).
\end{align}
This result is another example of how our construction goes beyond the (weakly coupled) result of the Boltzmann RTA. The $\sigma$-dependent sign coming from the CTA correction is negative, which complies with the simplest expectations derived from holography and perturbation theory \cite{Jeon:1994if,Kovtun:2004de,Buchel:2004di,Brigante:2007nu,Cremonini:2011iq,Grozdanov:2014kva,Grozdanov:2016zjj} that $\eta/s$ interpolates from a small to a large number between strong and weak coupling (which we associate with the RTA), respectively.

The correlators exhibit additional poles and a new branch cut as compared to the RTA cases discussed in \cite{Romatschke:2015gic,Bajec:2024jez}. By allowing us to access regions of the spectra at higher energies than in the Boltzmann RTA, this construction therefore elucidates how kinetic theory can approach known signatures of the spectra found by using other methods. In particular, additional cuts mimic the behavior of free thermal theories with an infinite sequence of branch cuts \cite{Hartnoll:2005ju}. In fact, we anticipate that the inclusion of each new level of the hierarchy will introduce a new cut into the spectrum, which is evident in the form of \eqref{hevolution}. This is because at every level, there will be an angular integration giving
\begin{align}
    \int \frac{d\Omega_n}{-i \omega + i v_a\cdot k_a+1/\tau_n}\sim \ln\frac{\omega-k+i/\tau_n}{\omega+k+i/\tau_n}.
\end{align}
The sequence of cuts is also reminiscent of the meromorphic ``Christmas tree'' structure known from holographic strongly coupled analyses \cite{Starinets:2002br,Kovtun:2005ev,Grozdanov:2018gfx}. Finally, the existence of new purely relaxing gapped poles also has analogy in holographic spectra at intermediate coupling \cite{Grozdanov:2016vgg,Grozdanov:2018gfx,Casalderrey-Solana:2018rle}.

\textbf{Future directions.---}Our work presents a novel quantitative method to extend the regularly used RTA Boltzmann equation to include higher levels of the BBGKY hierarchy using a physically motivated ansatz. In particular, the ansatz induces nontrivial correlations between constituents of a gas at scales beyond the region where the scattering occurs. This scheme enables the first analytic calculation of retarded correlators with effects due to the presence of higher BBGKY levels. Therefore, it opens the door for a quantitative kinetic discussion beyond the usual low-energy (infra-red) regime and also allows considerations of coupling constant corrections beyond the inherently weakly coupled, quasiparticle-based kinetic RTA description. 

Despite giving us analytic access to the BBGKY hierarchy, there remain a number of limitations in the present approach. The first is the choice of a simple, linear potential which leads to a coupling of the levels of the hierarchy. This could be relaxed in the future, but presumably at the cost of performing a numerical analysis. Moreover, the approximation scheme means that for each hierarchy level there is a corresponding relaxation timescale. Like transport coefficients in hydrodynamics, these timescales cannot be determined within the present theory and must be provided by other, microscopic means. Moreover, we took the relaxation times to be constant throughout this work, which while analytically tractable, is not realistic. An important extension of the present work would be to promote the relaxation time to be momentum dependent \cite{Kurkela:2017xis,Kurkela:2019kip,Rocha:2021zcw}, as well as to include temperature dependence in the string tension as determined from lattice simulations \cite{Kaczmarek:1999mm}.

There are a number of future directions that the present work inspires. It would be interesting to investigate more closely the connection between our results, various holographic computations in different physical states, as well as with the analytic structure of hydrodynamic correlators computed from an effective field theory \cite{Chen-Lin:2018kfl,Michailidis:2023mkd,Grozdanov:2024fle}. Moreover, we should also explore how the present story changes in an expanding Bjorken background, especially in the context of non-thermal fixed points \cite{Berges:2020fwq} and the late-time approach to the hydrodynamic attractor \cite{Soloviev:2021lhs,Jankowski:2023fdz}. Regardless of the precise physical question, we anticipate a number of applications of the CTA formalism to precision analyses and other extensions of the vast number of applications of the Boltzmann RTA across the physics literature. 

\textbf{Acknowledgements.---}We wish to thank Eduardo Grossi, David M{\" u}ller, Paul Romatschke, Andrei Starinets and Derek Teaney for fruitful discussions. The work of S.G. was supported by the STFC Ernest Rutherford Fellowship ST/T00388X/1. The work is also supported by the research programme P1-0402 and the project N1-0245 of Slovenian Research Agency (ARIS). A.S. was supported by funding from Horizon Europe research and innovation programme under the Marie Sk\l odowska-Curie grant agreement No.~101103006.

\appendix

\begin{widetext}
\section{Explicit expression of the correlation functions}\label{app:correlators}
Here, we state explicit expressions of all correlators computed from CTA at second order of the BBGKY truncation computed to $\mathcal{O}(\sig)$. First, the retarded current-current correlators are given in the longitudinal (charge density) channel by
\begin{align}
    G^{0,0}_J&= \frac{\chi  (\tone (L_2 \sig  \tone \chi  (-1+i \ttwo \omega )-2 i k (\tone-\ttwo))+L
   (1-i \tone \omega ) (\tone (\sig  \ttwo \chi -1)+\ttwo))}{-L \left(\sig  \tone^2 \chi
   -\tone+\ttwo\right)+\tone (L_2 \sig  \tone \chi +2 i k (\tone-\ttwo))},
\end{align}
and in the transverse channel by
\begin{align}   
   G^{1,1}_J&=
   \frac{\chi i \tone
   \omega}{4
   } 
   \left(\frac{2  (1-i\tone \omega)}{\tone^2 k^2}
   +\frac{(1-i \tone \omega)^2+\tone^2 k^2}{(i \tone k)^3} L\right)\nonumber\\
   &+\sig
   \frac{  \chi ^2 \omega  \left(\ttwo \left(k^2 L \tone^2 \ttwo+2 i k
   \tone (\tone-\ttwo)-L \ttwo (\tone \omega
   +i)^2\right)-L_2 \tone^2 \left(k^2 \ttwo^2-(\ttwo \omega
   +i)^2\right)\right)}{8 k^3 T_0 \tone \ttwo (\tone-\ttwo)},
\end{align}
where we introduced the following shorthand
\begin{align}
   L&=\ln\frac{\omega-k+\frac{i}{\tone}}{\omega+k+\frac{i}{\tone}}, \quad\quad
L_2=\ln\frac{\omega-k+\frac{i}{\ttwo}}{\omega+k+\frac{i}{\ttwo}}.\label{logs}
\end{align}
The remaining correlators follow from the Ward identities. 

The stress-energy tensor correlators are given in the longitudinal (sound, or spin-$0$) channel by 
\begin{align}
    \frac{G^{00,00}_T}{3(\varepsilon_0+P_0)}&=\frac{A+B\sig}{C+D\sig},
\end{align}
where
\begin{align}
    A&=4 k^3 \tone^2+k^2 L \tone (\tone \omega +2 i)+6 i k
   \tone \omega +3 i L \omega  (\tone \omega +i),\\
B&=i L L_2 (\tone-\ttwo) (\ttwo \omega +i)
   \left(\tone^2 \omega  \left(k^2 \ttwo^2+3 (\ttwo \omega
   +i)^2\right)+i \tone \left(k^2 \ttwo^2+3 \ttwo^2 \omega ^2+6 i
   \ttwo \omega -6\right)+3 i \ttwo\right)
   \nonumber\\
   &-2 k L \ttwo
   \Big(\tone^3 \omega  \left(2 i k^2 \ttwo^2-3 i (\ttwo \omega
   +i)^2\right)+\tone^2 \left(k^2 \ttwo^2 (-5-i \ttwo \omega )+3
   i \left(2 \ttwo^3 \omega ^3+5 i \ttwo^2 \omega ^2+\ttwo \omega
   +2 i\right)\right)
   \nonumber\\
   &+\tone \ttwo \left(4 k^2 \ttwo^2+3
   \ttwo^2 \omega ^2-18 i \ttwo \omega +9\right)+3 \ttwo^2 (-1+3
   i \ttwo \omega )\Big)
   \nonumber\\
   &+2 k L_2 \tone (1-i \ttwo \omega
   ) \left(-2 \tone^2 \left(k^2 \ttwo^2+3 (\ttwo \omega
   +i)^2\right)+\tone \ttwo \left(k^2 \ttwo^2+3 \ttwo^2
   \omega ^2+9 i \ttwo \omega -9\right)+3 \ttwo^2\right)\nonumber\\
   &-4 k^2 \tone \ttwo (\tone-\ttwo) \left(2 \tone \left(2 i k^2
   \ttwo^2-3 i (\ttwo \omega +i)^2\right)-3 \ttwo (3 \ttwo \omega +i)\right),\\
   C&=4 k^3 \tone^2+2 i k^2 L \tone+12 i k \tone \omega +6 i L \omega 
   (\tone \omega +i),\\
   D&=2 (\ttwo \omega +i) \Big(2 i k L \ttwo \left(\tone^2
   \left(k^2 \ttwo^2+3 \ttwo^2 \omega ^2+3 i \ttwo \omega
   +3\right)-6 \tone \ttwo+3 \ttwo^2\right)
   \nonumber\\
   &-2 i k L_2
   \tone \left(\tone^2 \left(k^2 \ttwo^2
   +3 (\ttwo
   \omega +i)^2\right)
   +3 \tone \ttwo (2-i \ttwo \omega )-3
   \ttwo^2\right)\nonumber\\
   &+12 k^2 \tone \ttwo (\tone-\ttwo) (-i \tone \ttwo \omega +\tone-\ttwo)-3 L
   L_2 (\tone-\ttwo)^2\Big)\nonumber\\
   &+4 i k \ttwo^3 (\tone-\ttwo) \left(i L \left(k^2 \tone+3 \omega  (\tone \omega
   +i)\right)+2 k \tone \left(k^2 \tone+3 i \omega \right)\right),
\end{align}
in the transverse (shear, or spin-$1$) channel by 
\begin{align}
\frac{G^{01,01}_T}{\varepsilon_0+P_0}&=\frac{a+b\sig}{c+d\sig},
\end{align}
where
\begin{align}
    a&=2 k \tone \left(3 (\tone \omega
   +i)^2-2 k^2 \tone^2\right)-3 L (\tone \omega +i) \left(k^2 \tone^2-(\tone \omega +i)^2\right),\\
b&=9 T_0^2 \tone \Big[i L \left((\tone \omega +i)^2-k^2 \tone^2\right) \Big(3 L_2 \omega 
   (\tone-\ttwo) (\ttwo-\tone)
   \left(-k^2 \ttwo^2+(\ttwo \omega +i)^2\right) \nonumber \\&-2 k
   \ttwo (\ttwo \omega +i) \left(3 \omega  (\tone-\ttwo)^2
   -2 i k^2 \tone \ttwo^2\right)\Big)\nonumber \\
   &-2 k \tone \Big(L_2 \left(k^2 \ttwo^2-(\ttwo \omega +i)^2\right) \left(2 k^2 \tone^2
   \left(\ttwo^2 \omega +i \tone\right)+3 \omega 
   (\tone-\ttwo)^2 (1-i \tone \omega )\right)\nonumber \\
   &-2 i
   k \ttwo (\tone-\ttwo) (\ttwo \omega
   +i) \left(2 k^2 \tone (\tone+\ttwo)-3 \omega 
   (\tone-\ttwo) (\tone \omega
   +i)\right)\Big)\Big],\\
   c&=3 i L \left(k^2 \tone^2-(\tone \omega +i)^2\right)+2 k
   \tone \left(2 k^2 \tone^2-3 i \tone \omega
   +3\right),\\
   d&=-9 T_0^2 \tone^2 \Big(i L \ttwo^3 \left(k^2
   \tone^2-(\tone \omega +i)^2\right)
   -i L_2
   \tone^3 \left(k^2 \ttwo^2-(\ttwo \omega
   +i)^2\right)\nonumber\\
   &-2 k \tone \ttwo (\tone-\ttwo) (-i \tone \ttwo \omega +\tone+\ttwo)\Big),
\end{align}
and in the tensor (or spin-$2$) channel by 
\begin{align}
\frac{G^{12,12}_T}{\varepsilon+P_0}&=\frac{3i\tone\omega }{16} \left[\frac{10}{3}\frac{1-i \tone \omega }{\tone^2 k^2 }
+\frac{2(1-i \tone \omega)^3}{\tone^4 k^4}+i \frac{((1-i\tone \omega)^2+\tone^2k^2)^2}{\tone^5 k^5}L\right]\nonumber\\
        &-\sig \frac{3 T_0^2 \omega }{64 \pi
   ^2 k^5 \tone^3 \ttwo^3 (\ttwo-\tone)}\Big{[}
    3 L \ttwo^4 \left((\tone \omega +i)^2-k^2 \tone^2\right)^2
    -3 L_2 \tone^4
   \left((\ttwo \omega +i)^2-k^2 \ttwo^2\right)^2 \nonumber\\
   &+2 i k
   \tone \ttwo (\tone-\ttwo)
   \left(\tone^2 \left(5 k^2 \ttwo^2-9 \ttwo^2
   \omega ^2-9 i \ttwo \omega +3\right)+3 \tone \ttwo (1-3 i \ttwo \omega )+3 \ttwo^2\right)
        \Big{]}.
\end{align}
Note that the $\sig=0$ terms correspond exactly to the correlators computed in \cite{Romatschke:2015gic,Bajec:2024jez}. Again, all remaining correlators can be determined via the Ward identities.

\end{widetext}

\bibliography{bbgky}

\end{document}